Magneto Transport of high *TCR* (temperature coefficient of resistance) $La_{2/3}Ca_{1/3}MnO_3$: Ag Polycrystalline Composites

V.P.S. Awana[1,*], Rahul Tripathi[1], S. Balamurugan[2], H. Kishan[1] and E. Takayama-Muromachi[2]

[1] *Superconducivity and Cryogenics Division, National Physical Laboratory, Dr. K.S. Krishnan Marg, New Delhi –110012, India*

[2] *Advanced Nano Materials Laboratory, National Institute for Materials Science (NIMS), 1-1 Namiki, Tsukuba, Ibaraki, 305-0044, Japan*

We report the synthesis, (micro)structural, magneto-transport and magnetization of polycrystalline $La_{2/3}Ca_{1/3}MnO_3$:$Ag_x$ composites with x = 0.0, 0.1, 0.2, 0.3 and 0.4. The temperature coefficient of resistance (*TCR*) near ferromagnetic (*FM*) transition is increased significantly with addition of Ag. The *FM* transition temperature ($T^{FM}$) is also increased slightly with Ag addition. Magneto-transport measurements revealed that magneto-resistance *MR* is found to be maximum near $T^{FM}$. Further the increased *MR* of up to 60% is seen above 300 K for higher silver added samples in an applied field of 7 Tesla. Sharp *TCR* is seen near $T^{FM}$ with highest value of up to 15 % for Ag (0.4) sample, which is an order of magnitude higher than as for present pristine sample and best value yet reported for any polycrystalline LCMO compound. Increased *TCR*, $T^{FM}$ and significant above room temperature *MR* of $La_{2/3}Ca_{1/3}MnO_3$:$Ag_x$ composites is explained on the basis of improved grains size and connectivity with silver addition in the matrix. Better coupled *FM* domains and nearly conducting grain boundaries give rise to improved physical properties of the $La_{2/3}Ca_{1/3}MnO_3$ manganites.

[*] Corresponding Author- e-mail: awana@mail.nplindia.ernet.in
Fax No. 0091-11-25726938

# INTRODUCTION

Mangenites with formula $RE_{1-x}A_xMnO_3$ (RE = trivalent rare earth and A = divalent alkali metal viz. Ca, Ba, Sr, Pb or Bi etc.) were widely studied for their unusual magneto-transport properties, please see couple of review and references there in [1-3]. These materials gave rise to huge magneto-resistance (*MR*) just below room temperature. For any practical viability, it is important to realize the *MR* at above 300 K (room temperature). In particular they can be used as the magnetic switch devices. Magnetically these compounds undergo paramagnetic (*PM*) to ferromagnetic (*FM*) transition coupled with near insulator to metal change over. The *MR* is basically seen at metal insulator transition temperature ($T^{MI}$), which is close to the para-*FM* transition temperature ($T^{FM}$). Not only the higher *T*, but the sharpness of $T^{FM}$ and $T^{MI}$ are desirable for sharp high *MR*. The sharpness of $T^{MI}$ is in general judged from temperature coefficient of resistance (*TCR*). Further, the sharper $T^{MI}$ is generally seen in vicinity of sharp *PM* to *FM* transition.

The ultimate aim is to get sharp $T^{MI}$/higher *TCR* at elevated temperature i.e., above room temperature, without reducing the *MR*. In these regards there had been some trials in literature [4-6]. For example by increasing the amount of $Mn^{4+}$ through substitutions like RE/Ag [4,5], or increase in overall oxygen content [1,7], one could shift the $T^{MI}$ and $T^{FM}$ above/close to the room temperature. However one needs not only the room temperature $T^{MI}$ and $T^{FM}$ but sharper *MI* and *FM* transitions for meaningful *MR* at these temperature. Sharpness of *MI* transition is defined by the *TCR*. Hence in general one needs to employ an strategy, where by both higher content of $Mn^{4+}$ and reasonable *TCR* could be achieved. The former guarantees the above room temperature $T^{MI}$ and the later high *MR*. The highest *TCR* values yet obtained were around 20% for Ag implanted LCMO laser ablated thin films [6,8] and 50% for $La_{2/3}(Ca,Pb)_{1/3}MnO_3$ single crystals [9]. For polycrystalline LCMO or similar compounds the *TCR* values yet seen had always been below 10%. Also important is not only to observe high *TCR* but it should be like a sharp peak with temperature, with in say 2-5 K for practical bolometers. In principle small temperature variation should be able to generate a significant drop across the bolometer [10]. For large applications, in certain cases one might require bulk large area rod like high *TCR* material instead of thin films or single crystals. Hence the high *TCR* polycrystalline large LCMO composites were warranted. In this direction we thought of improving the grains connectivity of the bulk mangenite material for better *TCR* by Ag additions to the widely studied mangenites La$_{2/3}$Ca$_{1/3}$MnO$_3$ (LCMO). The Ag additions in HTSC (*High T$_c$ cuprate superconductors*) compounds had proved to be very useful in terms of the improved grain boundaries [11]. In this article, we report the results of magnetization and magneto-transport for LCMO:Ag$_x$ composites with x = 0.0, 0.1, 0.2, 0.3 and 0.4. We found both increased *TCR*, $T^{FM}$ and significant above room temperature *MR* for LCMO:Ag$_x$ composites. The scanning electron micrograph (SEM) pictures of the LCMO:Ag$_x$ composites showed remarkable improvement in the grains morphology with Ag addition. Our results have direct implications towards the practical use of magneto-resistive mangenites.

EXPERIMENTAL

La$_{2/3}$Ca$_{1/3}$MnO$_3$ (LCMO): Ag composites were synthesized by solid-state reaction route with ingredients La$_2$O$_3$, CaCO$_3$, Mn$_2$O$_3$ and metallic Ag. The mixed powders were calcined at 1000 °C, 1100 °C and 1200 °C in air for 24 hours each. Finally the palletized ceramics were annealed in air for 48 hours at 1300 °C. For further loading of oxygen the final pellets were annealed in flow of oxygen at 800 °C for 48 hours and subsequently slow cooled to room temperature over a span of 12 hours. The structure and phase purity of the La$_{2/3}$Ca$_{1/3}$MnO$_3$ (LCMO): Ag composites were checked by powder X–ray diffraction (XRD) performed on a diffractometer (Rigaku RINT2200HF-Ultima) using the Ni–filtered Cu K$_\alpha$ radiation at 40 kV and 50 mA. The data were obtained between 20 and 80° 2θ in steps of 0.02°. Magnetization measurements were carried out in a commercial magnetometer with the superconducting quantum interference device (Quantum Design: MPMS-5S) between 5 K and 400 K under the applied field of 10 kOe. An isothermal magnetization curves were obtained with applied fields up to ±50 kOe at fixed (different) temperatures (T = 5, 50, 100, 300, 350 K). The transport and magneto-transport measurements were carried out in a commercial apparatus (PPMS–6600, Quantum Design) between 5 and 400 K in magnetic fields up to 70 kOe. SEM studies were carried out on these samples using a Leo 440 (Oxford Microscopy: UK) instrument.

RESULTS

Fig. 1 depicts the room temperature x-ray diffraction (XRD) patterns of for La$_{2/3}$Ca$_{1/3}$MnO$_3$ (LCMO): Ag composites. All the samples crystallize in orthorhombic perovskite structure with space group *Pbnm* having $a \sim b \sim a_p\sqrt{2}$ and $c \sim 2a_p$. For pure x = 0.0 sample, the lattice parameters are $a$ = 5.51 Å, $b$ = 5.47 Å and $c$ = 7.75 Å. This is in general agreement with previous reports for LCMO compounds [1-3]. Both the lattice parameters and the structural space group of the pristine sample remain invariant with successive addition of Ag in La$_{2/3}$Ca$_{1/3}$MnO$_3$ (LCMO): Ag composites. This is in contrast to La$_{2/3-x}$Ag$_x$Ca$_{1/3}$MnO$_3$ compounds, where substitution of Ag at La-site induces an orthorhombic to rhombohedral phase transformation [12,13]. For higher Ag content samples Ag is also seen within LCMO matrix main phase. It seems in our La$_{2/3}$Ca$_{1/3}$MnO$_3$ (LCMO): Ag composites Ag is not substituted into main LCMO lattice but remain rather as an additive in the system.

Fig. 2 (a) and (b) shows the SEM pictures of the pristine and Ag (0.4) samples, in same magnification of 10μm for inter comparison. It is seen that grain morphology is improved tremendously with inclusion of Ag in the matrix of pristine LCMO.

Normalized resistance ($R_T/R_{400}$) versus temperature plots down to 5 K for La$_{2/3}$Ca$_{1/3}$MnO$_3$ (LCMO): Ag composites are depicted in Fig. 3. Qualitatively all the samples are (a) semiconducting down to around 280, (b) passes through a insulator-metal transition ($T^{MI}$) below 280 K, and (c) exhibits negative upturn at below 50 K for pristine and low (x > 0.2) Ag content samples. The absolute value of resistivity at 400 K decreases with an increase in x. Though the $T^{MI}$ occurs below 280 K for all the samples the sharpness of the transition is improved significantly with addition of Ag. For example for pure LCMO the metallic transition is broad and instead look like a hump over a range of 280 K down to 100 K. It seems the pristine sample needs further improvement. However within present constraints as we discuss later within same heat treatments the LCMO:Ag composites proved to be of very high quality, so we leave the pristine compound as such in the present study. A very recent study on pristine similar La$_{2/3}$Ba$_{1/3}$MnO$_3$ (LBMO) compound had once again shown that the quality of the sample changes dramatically with sintering temperature [14]. The sintering temperature of 1300 $^0$C must be good enough for LCMO:Ag composites but slightly less for pristine LCMO. We are working on this issue and the results will be reported later in a separate communication. Further for pristine sample an upturn in resistance is seen below 50 K, which is not a good sign for a homogenous ferromagnetic compound. With addition of Ag, the $T^{MI}$ becomes quite sharper and the below 50 K upturn is not seen for Ag content above 0.2. For Ag (0.3) and Ag (0.4) samples the *R/T* remains metallic from 280 K ($T^{MI}$) down to 5 K. The sharpness of $T^{MI}$ can be defined by *TCR*, where *TCR* is [1/*R* x (d*R* / d*T*] x 100. The *TCR* is around 1.2% for pristine sample and is 15% for Ag (0.4) sample. The *TCR* (*T*) plots for pure and Ag(0.4) LCMO compounds are depicted in inset of Fig. 3. The interesting point is that *TCR* is improved by an order of magnitude and is best value yet reported for any polycrystalline LCMO compound. Higher *TCR* of the present LCMO-Ag composites could be very useful for infra red/bolometric detector applications [10]. In case of Ag (0.1), though the most structure of pure LCMO hump is absent, the transition is seen in two steps. In case of Ag = 0.2, 0.3 and 0.4 samples the transition is quite sharp and the obvious LCMO polycrystalline hump is not seen, further *TCR* is improved by nearly an order of magnitude. To our knowledge these are first LCMO:Ag polycrystalline compounds having very sharp $T^{MI}$, which is generally the case for aligned LCMO films [6,8,15] or single crystals [9]. For large applications occasionally the bulk compounds in form of solid rods etc are required and hence the present LCMO:Ag composites will prove to be very useful in this direction.

The magnetization results of the $La_{2/3}Ca_{1/3}MnO_3$ (LCMO): Ag composites are shown in Fig. 4. The *M/T* plots being taken at 10 kOe are given in main panel for all samples and the *M/H* for *H* up to 5 T are given in two insets for pristine and Ag (0.1) samples. Mainly the $T_c$ (Para –ferro transition temperature) is seen below 300 K and the same remains invariant for all the samples. This is further seen from the fact the *M/H* is seen linear at 300 and 350 K for all the samples, please see the insets of Fig. 4 for pure and Ag(0.1). The only difference is seen for *M/T*, for example the saturation moment (emu/gram) is increased with an increase in x from 78 emu/gram for x = 0.0 to 105 emu/gram for Ag (0.4) sample. The unchanged $T_c$ with x is in accordance with invariant $T^{MI}$ (Fig.4) of LCMO: Ag composites.

Fig. 5(a) depicts the magneto transport of LCMO pure sample. The zero-field *R/T* plot is already discussed in Fig. 3, which is comprised of three parts, the insulator regime above 280 K, the metallic region between 280 K down to 50 K, and Kondo like low temperature upturn below 50 K. Under magnetic field of 7 Tesla, the below $T^{MI}$ (280 K) hump is reduced sharply exhibiting sufficient MR [$(R_0-R_H/R_0)$ x 100] below $T^{MI}$ down to 5 K. The below 50 K Kondo-like upturn of R/T is melted in 7 Tesla field and compound becomes metallic down to 5 K. We also performed the fixed temperature and varying field *MR* experiments on our pure sample at 300, 250, 200, 100 and 50 K, please see inset of Fig. 5 (a). The *MR* of up to 35% is seen in 7 Tesla field at 300 K. The shape of $MR^{300K}$ is U type and hence the low field *MR* is quite less (~1% at 1 Tesla). At lower temperatures of 250 K, 200K, 100 K and 50 K, the *MR* shape is V type indicating tunneling magneto-resistance in this regime. The maximum *MR* is seen at 250 K and least 300 K in nearly all applied fields.

The magneto transport of LCMO: Ag (0.1) is similar to that as for pristine LCMO, though with reduced width or sharper hump, and also similar *MR* characteristics (plots are not shown). For LCMO: Ag (0.2) the *R/T* plots in zero and 7 Tesla fields along with fix temperature *MR* are shown in Fig. 5 (b). The insulator-metal transition at $T^{MI}$ (280 K) is quite sharp with high *TCR* in comparison to pristine sample. The low temperature *R/T* upturn is not seen in this compound. Further the *MR* is seen right up to 400 K. The fixed temperature (300 K) *MR* in various applied fields is though U type similar to that as for pristine LCMO, but the *MR%* value is much higher. For example at 300 K in 7 Tesla field the *MR* of around 43% is achieved, which is the highest value than all other low *T* values. Fig. 5(c) and 5(d) depicts the magneto transport of LCMO: Ag (0.3) and LCMO: Ag (0.4) samples. Briefly, both these compounds have much sharper transitions with higher *TCR*, absence of low temperature, below 50 K *R/T* upturn and sufficient *MR* at 300 K. The sharpness of the transition near $T^{MI}$ (280 K) is comparable to aligned thin films [6,8,15] of LCMO, and to our knowledge is the best one yet obtained for any polycrystalline mangenites. As far fixed temperature and varying field *MR* is concerned its shape is changed from U type to V type. And is maximum at all fields up to 7 Tesla at 300 K. Infact with an increase in Ag content the low *T* below 300 K *MR%* decrease and the room temperature 300 K increase. For LCMO: Ag (0.4) sample the *MR* of up to 60% is observed in 7 Tesla field at 300 K. Even at low fields of say 1 Tesla the *MR* of up to 16% is seen.

The results of magnetization and magneto transport of LCMO:$Ag_x$ composites can be summarized as follows:

1. Ag does not substitute into host LCMO matrix in our polycrystalline LCMO:$Ag_x$ composites but improved significantly the grain morphology of the host.
2. The $T^{MI}$ and $T_c$ remain nearly invariant with increase in x (Ag content).
3. The sharpness of insulator-metal transition increases dramatically with increase in x (Ag content) and high (15%) sharp *TCR* is observed for Ag composites
4. Though the *MR* is seen in all the samples right from 300 K down to 5 K, the same is negligible at 300 K and more at lower *T* for pure sample and is maximum at 300 K and least at 5 K for Ag (0.4) containing compounds.

**DISCUSSION**

Now we try to explain in broad sense the main results summarized above, as far as (1) is concerned there are some very recent reports in literature, which deals with the LCMO composites viz. LCMO/ZnO [16], LCMO/CuO [17], LPbMO/$Fe_2O_3$ [18], LCMO/NiO [19] and LCMO/$ZrO_2$ [20]. In all these composites the quality of the LCMO was deteriorated. In case of LCMO:Ag there is a good probability of Ag being distributed at grain boundaries and hence providing better connectivity of grains, which is seen in Fig. 2(b). In most of other composites [16-20] the additives either got substituted in some quantity in the lattice or got segregated as clusters in the matrix of LCMO. This gave rise to the poor quality of LCMO composite.

As far as point (2) is concerned the $T^{MI}$ and $T_c$ remain nearly invariant with increase in x, primarily due to the fact that in our composites the La/Ag substitution [4,5], which could increase the effective amount of $Mn^{4+}$ and hence $T^{MI}$ and $T_c$ has not taken place. In case of Ag substitution at La-site, instead of remaining as additive in the matrix, though the $T^{MI}$ and $T_c$ increased but the electrical conduction was not improved significantly, in fact the same got detoriated in some cases [4,5]. In our opinion the Ag addition in LCMO could results in (a) as La-site substitution, (b) remain as clusters in the matrix and (c) distributes finely at grain boundaries of LCMO. As we told earlier (a) results in increased $T^{MI}$ and $T_c$ due to increase in $Mn^{4+}$ content, (b) neither affects $T^{MI}$ and $T_c$ nor does improve the conduction, and (c) though does not improve upon the $T^{MI}$ and $T_c$ but improves the grains morphology and the transport properties of the compound. This naturally answers point (3), where we see that in our composites the sharpness of metal-insulator transition increases dramatically due to better grains morphology/connectivity in LCMO: Ag composites, please see Fig. 2(b). As far as the process (a), (b) and (c) are concerned the best combination could be the simultaneous process (a) and (c), i.e., increase $T^{MI}$ and $T_c$ to sufficiently above room temperature by partial La site Ag substitution and also leave some Ag at grain boundaries of LCMO grains to improve upon the grains morphology/better connectivity/better coupled FM domains to increase the *TCR*/sharpness of *M-I* transition. The effectiveness of process (a), (b) or (c) depends upon the sample synthesis conditions. In our case we mixed the Ag with LCMO in calcinations process itself at 1000 $^0$C and went up to 1300 $^0$C in steps of 100 $^0$C with intermediate grindings/pulverization etc. Our synthesis process has resulted in process (c), i.e. the distribution of Ag at grains boundaries of LCMO and hence improved grains connectivity and sharper *M-I* transition. More efforts are required to both increase $T^{MI}$ and $T_c$ by process (a) and sharper *M-I* transition by process (c) with in the same sample with improved heating schedules.

Regarding point (4), the decrease in *MR* at low temperatures and increase at higher *T* with x in LCMO: Ag composites, we would like to stress that better connectivity of grains and sharper $T^{MI}$ is the reason behind the same. Sharper $T^{MI}$ guarantees the higher *TCR* and hence the *MR* process is in higher magnitude and at higher *T* in LCMO: Ag composites.

## SUMMARY AND CONCLUSION

Summarily we have synthesized bulk LCMO: Ag composites by solid state reaction route, which resulted in substantially improved transport properties in terms of sharp *M-I* transition and high *MR* near room temperature. The $T^{MI}$ and $T_c$ could not be increased above room temperature but the temperature coefficient of resistance (*TCR*) near room temperature is improved significantly. The below 50 K *R* (*T*) upturn is not seen, which is further a sign of improved electrical conduction in these composites. With addition of Ag the practically important higher *T MR* is improving, but the low *T MR* got diminished. Unlike some previous reports in our case, the Ag is not getting substituted at La –site but remaining at grain boundaries and hence improving upon significantly the electrical transport via better grains connectivity.

## ACKNOWLEDGEMENT

Authors from NPL appreciate the interest and advice of Prof. Vikram Kumar (Director) NPL in the present work. One of us Rahul Tripathi acknowledges the financial help in form of JRF-NET fellowship from the UGC-India.

**FIGURE CAPTIONS**

Figure 1. X-ray diffraction patterns of LCMO: Ag composites at room temperature.

Figure 2. SEM pictures of pure LCMO and LCMO: Ag(0.4) samples at same magnification.

Figure 3. $R(T)$ plots of LCMO: Ag composites, the inset shows the *TCR%* for pure and Ag (0.4) samples.

Figure 4. Magnetization –Temperature ($M/T$) plots for LCMO: Ag composites, the insets show the $M/H$ for pristine and Ag(0.1) samples.

Figure 5 (a). $R(T)$ plots of pristine LCMO in 0 and 7 Tesla applied fields, the inset shows the MR at various $T$ in applied fields of $\pm 7$ Tesla.

Figure 5 (b). $R(T)$ plots of LCMO (Ag0.2) in 0 and 7 Tesla applied fields, the inset shows the MR at various $T$ in applied fields of $\pm 7$ Tesla.

Figure 5 (c). $R(T)$ plots of LCMO (Ag0.3) in 0 and 7 Tesla applied fields, the inset shows the MR at various $T$ in applied fields of $\pm 7$ Tesla.

Figure 5 (d). $R(T)$ plots of LCMO (Ag0.4) in 0 and 7 Tesla applied fields, the inset shows the MR at various $T$ in applied fields of $\pm 7$ Tesla.

FIG. 1 XRD of LCMO: Ag$_x$

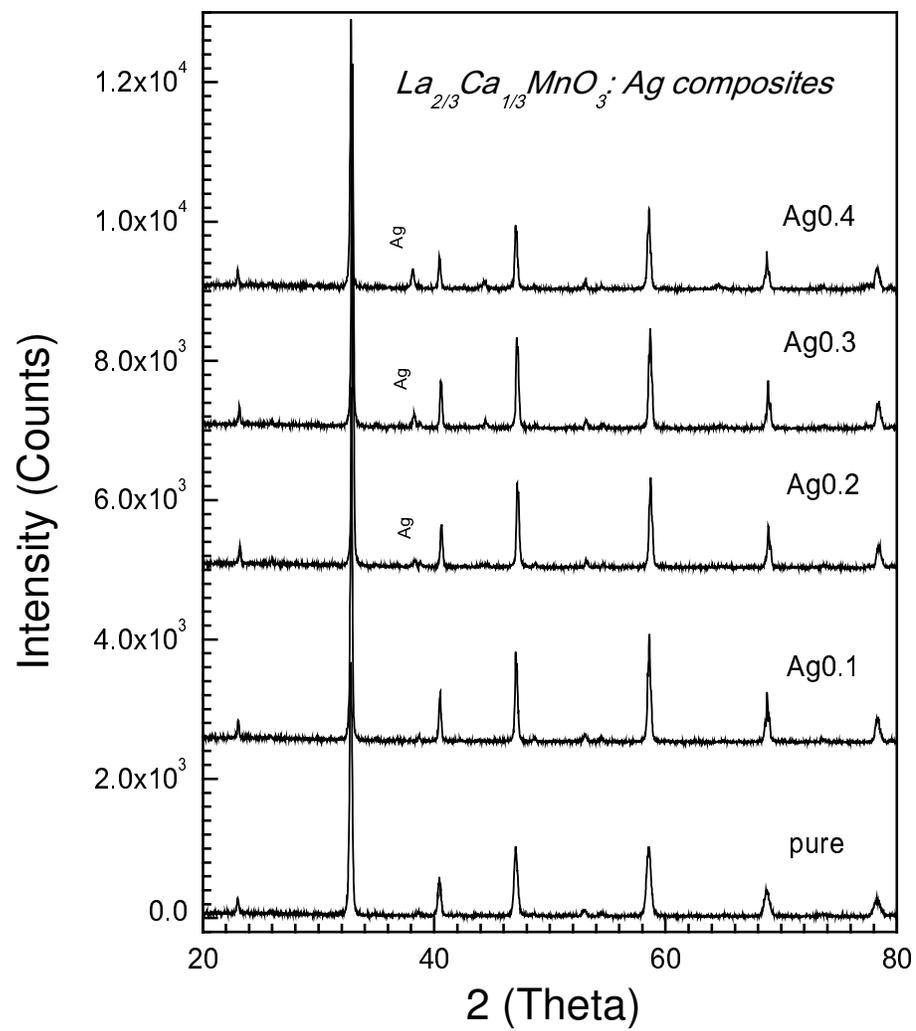

FIG. 2(a) SEM of pure LCMO

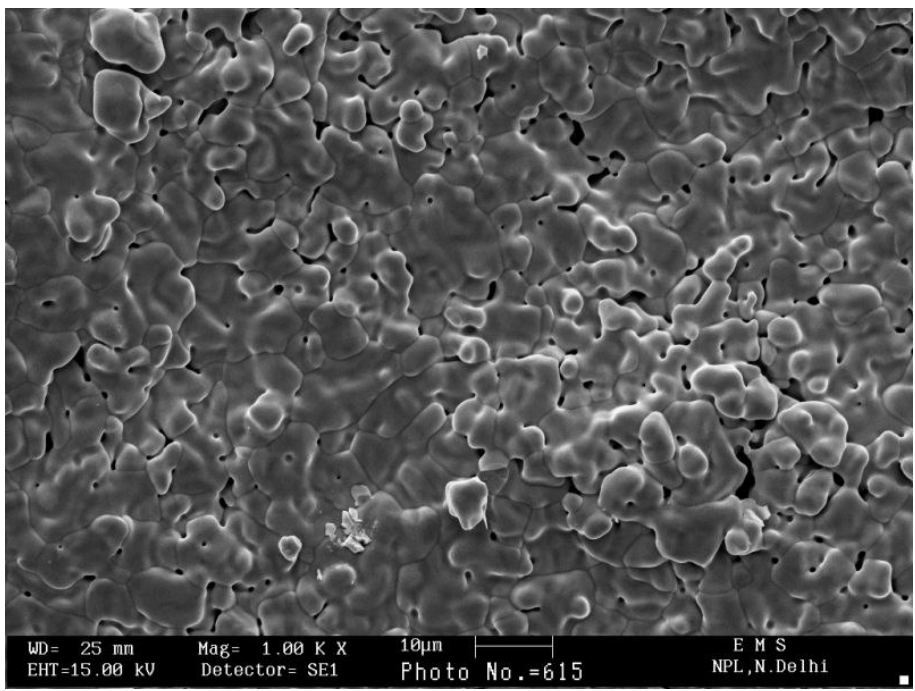

FIG. 2(b) SEM of LCMO (Ag0.4)

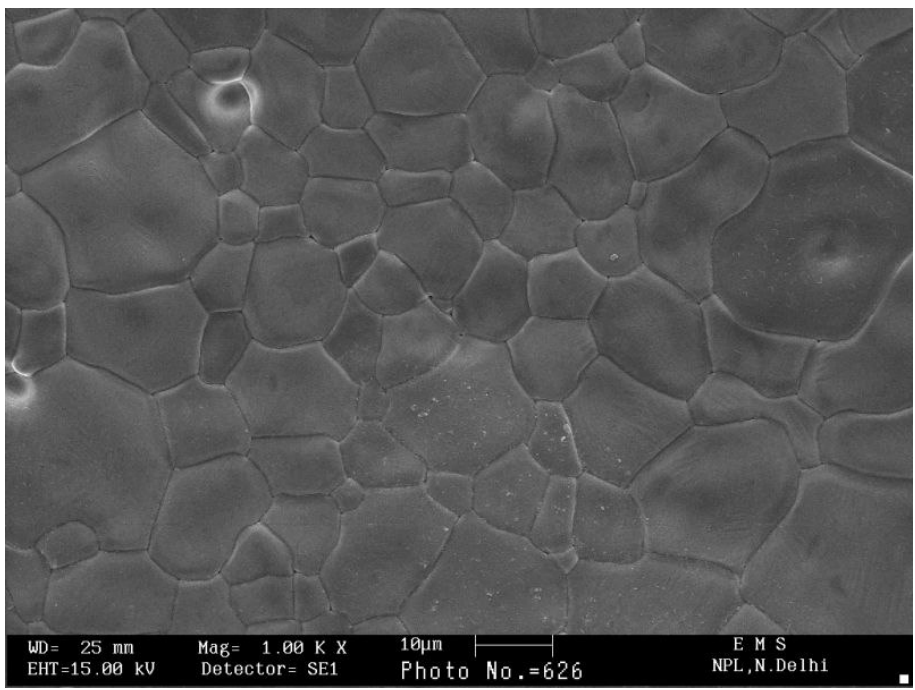

FIG.3 $R/T$ of LCMO: $Ag_x$ composites

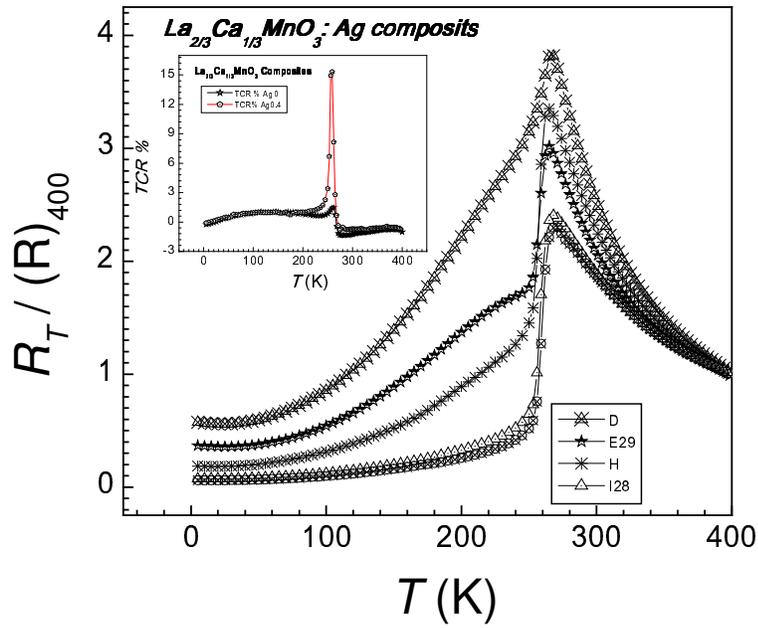

FIG.4 $M/T$ of LCMO: $Ag_x$

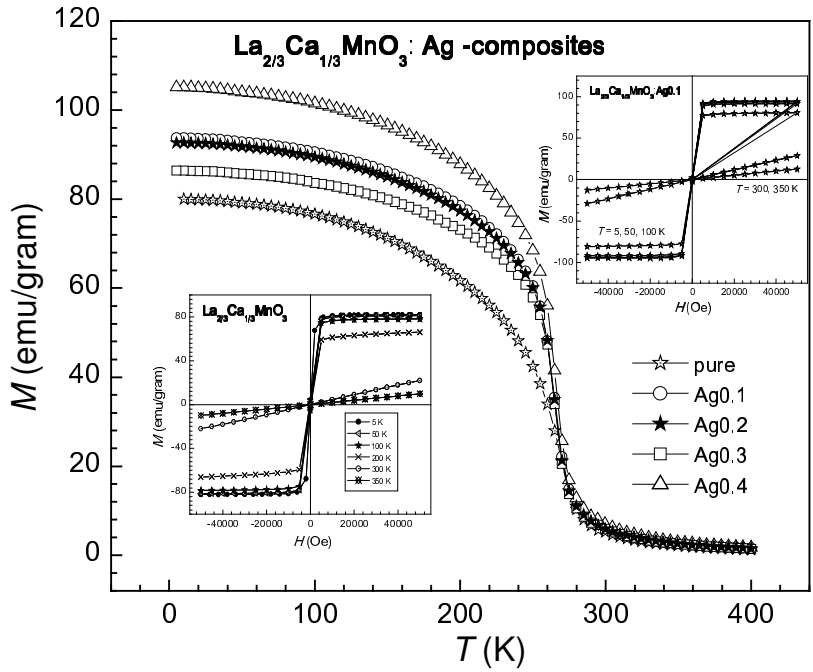

FIG.5 (a) *MR* of LCMO

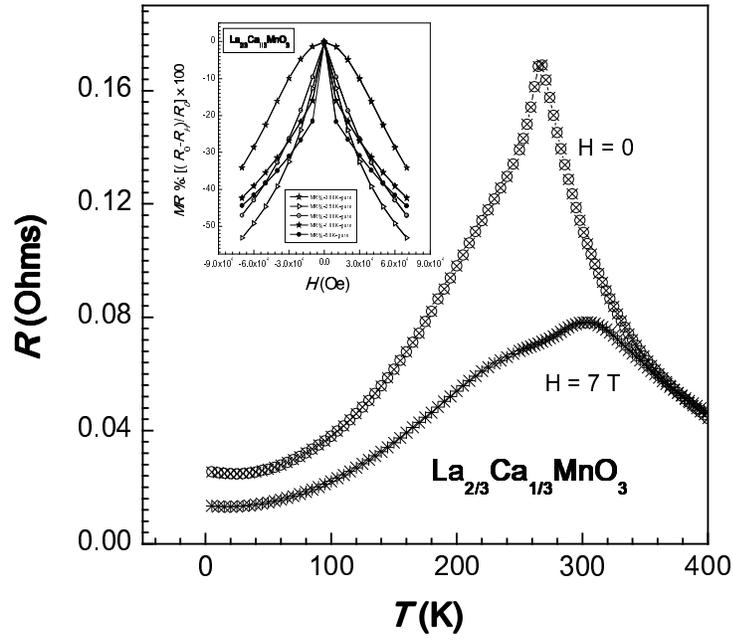

FIG.5 (b) *MR* of LCMO: Ag (0.2)

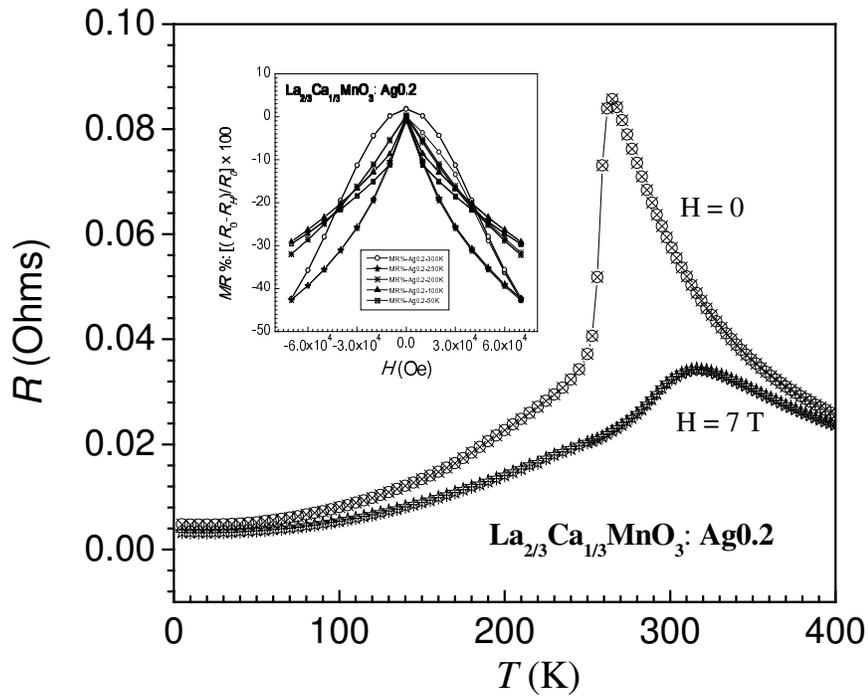

FIG. 5(c) *MR* of LCMO: Ag (0.3)

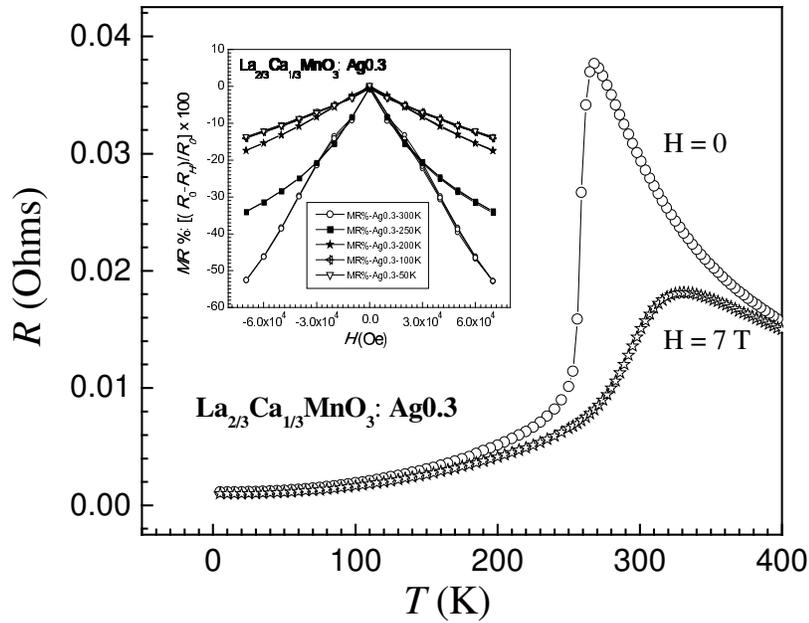

FIG. 5(d) *MR* of LCMO: Ag (0.4)

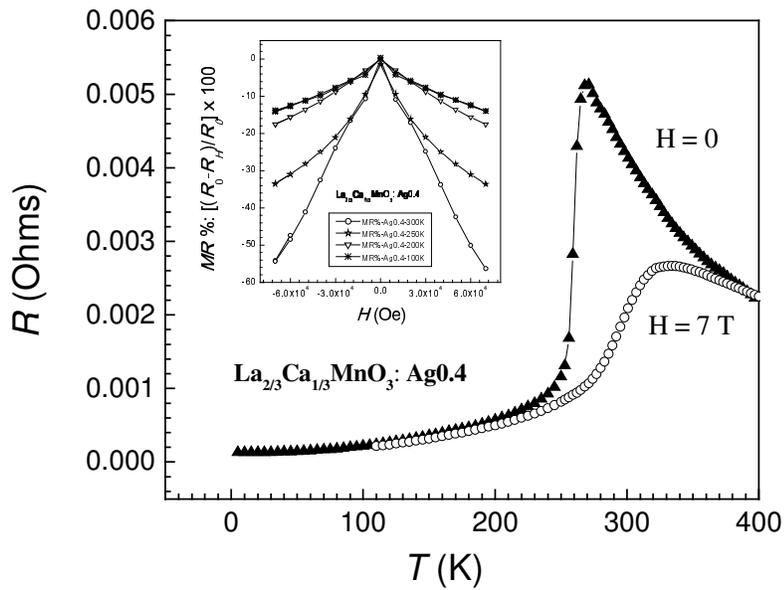